\documentclass[twocolumn,showpacs,preprintnumbers,amssymb]{revtex4}
\usepackage{epsfig}

\usepackage{afterpage}

\begin{document}

\title{Transport properties of microstructured ultrathin films of
La$_{0.67}$Ca$_{0.33}$MnO$_3$ on SrTiO$_3$}
\author{C. Beekman, I. Komissarov, M. Hesselberth and J. Aarts}
\affiliation{Kamerlingh Onnes Laboratory, Leiden University, P.O. box 9504, 2300RA
Leiden, The Netherlands}
\date{\today}

\begin{abstract}
We have investigated the electrical transport properties of 8~nm thick
La$_{0.67}$Ca$_{0.33}$MnO$_3$ films, sputter-deposited on SrTiO$_3$ (STO), and
etched into 5~$\mu$m-wide bridges by Ar-ion etching. We find that even slight
overetching of the film leads to conductance of the STO substrate, and asymmetric
and non-linear current-voltage (I-V) characteristics. However, a brief oxygen plasma
etch allows full recovery of the insulating character of the substrate. The I-V
characteristics of the bridges are then fully linear over a large range of current
densities. We find colossal magnetoresistance properties typical for strained LCMO
on STO but no signature of non-linear effects (so-called electroresistance)
connected to electronic inhomogeneites. In the metallic state below 150~K, the
highest current densities lead to heating effects and non-linear I-V
characteristics.
\end{abstract}

 \maketitle

Doped manganese oxides such as La$_{1-x}$Ca$_x$MnO$_3$ are of interest since, in a
certain range of doping, a combined insulator-to-metal and
paramagnetic-to-ferromagnetic transition can take place. One consequence is the
well-known Colossal Magnetoresistance effect, but another is the fundamentally
interesting phenomenon of phase separation. The susceptibility of the phase
transition to chemical and crystallographic disorder (doping disorder, oxygen
non-stoichiometry, defects from strain relaxation, twinning, grain boundaries) can
lead to an inhomogeneous state in which the insulating and metallic phases coexist
on a variety of length scales. In such systems, the percolative nature of the
conductance may lead to strongly non-linear behavior and a large sensitivity to
electric fields, which can be useful for a variety of applications. Lately,
therefore, there has been renewed focus on conductance issues, leading to various
different observations. Non-linearities, presented as a strongly decreasing
resistance as function of increasing current density, were reportedly measured on
microbridges made from films of La$_{0.7}$Ca$_{0.3}$MnO$_3$ and
La$_{0.85}$Ba$_{0.15}$MnO$_3$ grown on STO \cite{gao03}. Similar observations were
reported on samples made with La$_{0.7}$Ca$_{0.3}$MnO$_3$ \cite{zhao05}. In both
cases it was suggested that these so-called electroresistance (ER) effects are due
to phase separation. In other experiments, microbridges were subjected to high
currents ("current processing"), and non-linear as well as asymmetric
current-voltage (I-V) characteristics were subsequently found in the two-point
resistance \cite{sun05,xie06}. This was tentatively ascribed to the formation of
junction-like structures in the films, and therefore intrinsic, although
modification of the interface between the metal electrodes and the oxide film by the
current was not fully ruled out. The interface is a known complication in 2-point
geometries; non-linear and asymmetric I-V characteristics were demonstrated in
rectifying Ti/Pr$_{0.7}$Ca$_{0.3}$MnO$_3$ contacts \cite{sawa04}, in a $p$-$n$
heterostructure involving La$_{0.7}$Ca$_{0.3}$MnO$_3$ and Nb-doped SrTiO$_3$ (STO)
\cite{zhang05}, and in Ag-La$_{0.7}$Ca$_{0.3}$MnO$_3$ heterostructures
\cite{shang06}. However, 4-point measurements on La$_{0.8}$Ca$_{0.2}$MnO$_3$
microbridges also showed current-induced ER \cite{hu04,hu05}, and it was concluded
that high currents can change the balance in the coexistence of the different
phases. All of the above microbridges are still relatively large, with typical film
thicknesses of 100~nm and bridge widths around 50~$\mu$m. Phase separation phenomena
may be found down to very small length scales, in particular when strain and strain
relaxation also play a role \cite{faeth99,biswas00}. The question then arises
whether similar ER effects can be seen in smaller bridges and thinner films. Here we
note that special care has to be taken in the structuring. The commonly used
Ar$^+$-etching technique easily damages the STO substrate, which results in a
conducting surface layer after etching \cite{kan05}. Current leakage through this
layer interferes with the transport measurements and intrinsic current effects will
be obscured. This problem can be
overcome by a brief oxygen plasma etch, as will be shown below. \\
\indent Epitaxial films of La$_{0.67}$Ca$_{0.33}$MnO$_3$ with a typical thickness of
8~nm were grown on (001)STO substrates by DC sputtering in an oxygen pressure of
300~Pa, at a growth temperature of 840$^{\circ}$C. The substrate surface was treated
to have single termination of TiO$_2$, and had a misorientation of 1$^{\circ}$
towards [010] in order to improve the smoothness of the film. A resist mask was
patterned by e-beam lithography to yield a structure for 4-point measurements, with
a bridge width of 5~$\mu$m, a distance between the voltage contacts of typically
16~$\mu$m and the orientation of the bridge perpendicular with respect to the step
edges of the substrate. The film was etched with Ar-ions (beam current: 10 mA, beam
voltage: 350 V) during 40~s; with a calibrated etch rate 0.31~nm/s, this means
overetching of 14~s, in order to be certain to remove all of the film. Temperature
and magnetic field regulation were done in a Physical Property Measurement System
(PPMS: Quantum Design) but external current sources and nanovoltmeters were used to
perform most of
the transport measurements. \\
\begin{figure}[!h]
\includegraphics[width=9cm,angle=0]{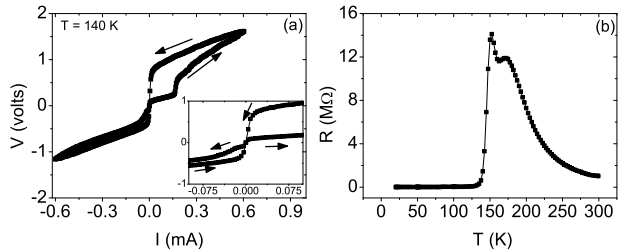}
\caption{\footnotesize{(a) I-V curve at T = 140 K of an LCMO bridge on STO after 14
sec overetching; insert : behavior near zero current); (b) Resistance measured as
averaged voltage (see text) versus T of the same sample.}}\label{fig1}
\end{figure}
\indent I-V characteristics were measured on the microbridges in the temperature
range of 10~K - 300~K using currents up to 0.6~mA, which corresponds to a current
density $J$ = 1.5$\times 10^{10}$ A/m$^2$. A typical one, taken at 140~K, is shown
in Fig.\ref{fig1}a. It shows nonlinearity and a large asymmetry between opposite
current directions, as well as hysteresis at high $J$. Similar curves could be
observed at all temperatures. Still, by simply averaging voltages at small positive
and negative currents (0.1~$\mu$A; this was performed with the PPMS electronics) the
'resistance' $R$ (Fig.\ref{fig1}b) shows a sharp phase transition at 150 K, as
expected for films under tensile strain \cite{aarts98}. Apparently, the measurements
at least partly probe the bridge structure, but since this is a 4-point measurement,
the asymmetry indicates that not all current is flowing between the voltage
contacts. It is known that Ar etching of STO causes the formation of a conducting
surface layer by the removal of oxygen \cite{kan05}. To  conclusively show this in
our system, we etched a $w$~= 1.5~mm wide STO substrate for 30 sec. After etching, 4
Au/MoGe contacts \cite{note1} where sputtered on top ($\ell$ = 2.6 mm between the
voltage contacts). The I-V curves (Fig.\ref{fig2}) are symmetric down to 150 K,
although below this temperature they become slightly nonlinear. Fig.\ref{fig2} also
shows the T-dependent sheet resistance $R_{\square} = R w / \ell$ obtained at low
bias. The dependence is metallic, $R_{\square}$ is very close to the number reported
in ref.~\cite{kan05}, but also important to note is that the absolute value of $R$
is in the range of (only) 1~k$\Omega$ - 10~k$\Omega$, which makes it substantially
smaller than the expected peak value for $R$ in our microbridge, or even for $R$ in
significantly wider and thicker bridges. \\
\begin{figure}[!h]
\includegraphics[width=8cm,height=4cm,angle=0]{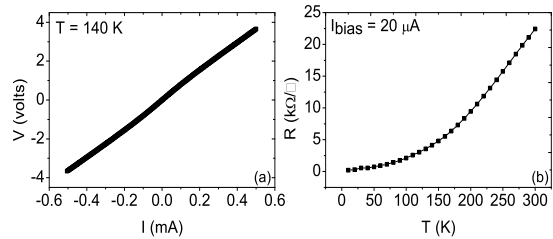}
\caption{\footnotesize{(a) I-V curve of 30 sec Ar-etched STO (1.5 x 2.6 mm$^2$) at T
= 140 K; (b) R-T plot of the same sample obtained from the I-V curves at 20 $\mu$A.
}}\label{fig2}
\end{figure}
\indent In order to restore the insulating properties of the substrate,
post-annealing in an oxygen environment would probably be possible, but this is
often unwanted as a process step; in our case, it might lead to strain relaxation
and additional defects in the film. Instead, after the Ar-etch we subjected
structures with the resist still in place to an O$_2$ plasma for typically
1~-~2~min. and found that the substrate had become insulating again while resist was
still present. Generally, the procedure appears to work for small amount of etching
of the STO. Substrates which were Ar-etched for more than 1 min. did not show
recovery anymore, even after 4~min. of plasma treatment (by which time the resist
layer had been removed). We surmise that oxygen loss can be recovered by the plasma,
but that more structural damage to the STO (amorphisation) renders this impossible.

\begin{figure}[h]
\includegraphics[width=8cm,height=4cm,angle=0]{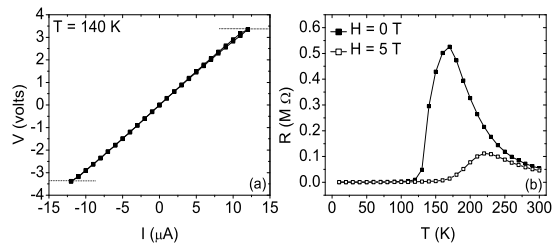}
\caption{\footnotesize{(a) I-V curve at T = 140 K of an LCMO bridge on STO after 4
sec overetching and O$_2$-plasma treatment. The thin lines indicate the voltage
limit for the measurement; (b) R-T plot at ~magnetic field B = 0~T and 5~T of the
same sample.}}\label{fig3}
\end{figure}

In Fig.~\ref{fig3} we show an I-V curve and the R(T) plot for a microbridge which
was overetched by 4 sec and then plasma-treated. The result clearly shows that when
the STO substrate is restored to its insulating state, the microstructured LCMO thin
film has linear and symmetric I-V characteristics in four point measurements. No
electroresistance is observed in our LCMO bridges.
Note that the current densities we used, between 2.5x10$^7$ A/m$^{2}$ and
1.5x10$^{10}$ A/m$^{2}$, lie within the range for $J$ where large resistance
variations were reported in refs.~\cite{gao03,zhao05}. In those cases no details are
given about I-V characteristics or the effects of microstructuring, but the samples
are different from ours, since they are typically 100~nm thick. A possible
explanation for the quite strong discrepancies is that our films are very
homogeneous even on submicrometer scales, in particular since strain relaxation has
not yet set in. It seems probable that the grain structure and the disorder in the
films determine possible ER effects to a large degree, as was surmised in
ref~\cite{zhao05}.
\begin{figure}[!h]
\includegraphics[width=8cm,height=4cm,angle=0]{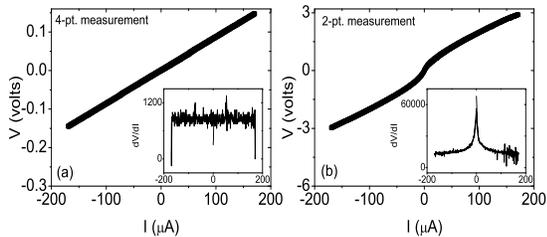}
\caption{\footnotesize{I-V curves of an LCMO bridge on STO at T = 10 K (a) in
4-point configuration;  (b) in 2-point configuration. Insets show derivatives
d$V$/d$I$. }}\label{fig4}
\end{figure}
\\
\indent To observe possible effects of the contacts we also measured the treated
sample in a two-probe configuration, with current injected through the voltage pads.
The results are shown in Fig.~\ref{fig4} where we compare the I-V curves for two and
four point measurements taken at 10~K. The 4-point measurement shows a linear and
symmetric IV curve as expected. The 2-point resistance is significantly larger
(around a factor of 5 after correction for the extra lead resistance in the contact
pads). This can be attributed to a large contact resistance. Another feature is the
nonlinearity of the I-V curve, with the corresponding peak in the derivative dV/dI
(see inset) clearly visible and most probably caused by the presence of a
barrier at the contact-film interface. \\
\begin{figure}[t]
\includegraphics[width=8cm,angle=0]{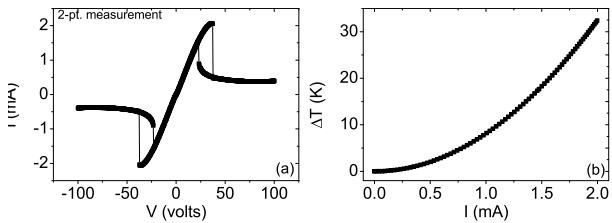}
\caption{\footnotesize{a)I-V curve of an LCMO bridge on STO at T = 50 K measured in
a 2-point configuration. b) Calculation of the expected Joule heating (see text)
represented as temperature increase $\Delta$T versus applied current I.
}}\label{fig5}
\end{figure}

\indent In the 2-point measurements in the metallic state we also found nonlinear
effects in the I-V characteristics which we ascribe to Joule heating. A measurement
at 50~K is shown in Fig.\ref{fig5}$a$ where the voltage-driven system suddenly
switches to a lower current (higher resistance). It is not straightforward to make
an estimate of the effect. The measured resistance is dominated by the contacts, but
the area of the contacts is much larger than the bridge so that it is not a priori
clear in which part of the structure the heating occurs. If we still assume it is
the bridge, we can estimate the temperature increase $\Delta T$ from a current $I$
using the following equation taken from \cite{padhan04}, $ \Delta T (T, I) =
(2I^{2}\rho (T + \Delta T))/(S \kappa_{sub}(T + \Delta T) $, with $\rho$(T) the
specific resistance of the bridge at temperature $T$, $S$ its cross-section, and
$\kappa$ the thermal conductance of the substrate. Taking $\rho$(T) in the metallic
state independent of temperature and use the value $\rho$~= 260~$\mu\Omega cm$ found
in the 4-point measurement, then with $\kappa$ = 16 WK$^{-1}$m$^{-1}$
\cite{steig68}, we find $\Delta$(T) $\approx$ 30~K at 2~mA (Fig.\ref{fig5}$b$). The
model is quite crude, but the result at least indicates that heating effects cannot
be negelected in our bridges. The mechanism leading to the switching behavior is not
yet understood, however, and needs further investigation.

\indent In summary we conclude that the observed peculiarities in the I-V
characteristics of our films are caused by the Ar-etching and are not an intrinsic
feature. We can restore the insulating STO surface layer by an O$_2$ plasma
treatment and then find no electroresistance effects in our thin and homogeneous
films. We further demonstrated that contact resistance and Joule heating can
introduce non-linear effects which are not intrinsic to the material under study. \\
\indent This research is in part supported by NanoNed, a nationale nanotechnology
program coordinated by the Dutch Ministry of Economic Affairs", and in part by the
'Stichting voor Fundamenteel Onderzoek der Materie (FOM)', which is financially
supported by the 'Nederlandse Organisatie voor Wetenschappelijk Onderzoek (NWO)'
%

\bibliographystyle{amsplain}

\begin{thebibliography}{99}
\bibitem{note1} amorphous Mo$_{70}$Ge$_{30}$ was found to be a good adhesion layer
for the Au contact.
\end{thebibliography}

\end{document}